\documentclass[twocolumn]{aastex631}

\usepackage{savesym}
\savesymbol{tablenum}
\usepackage{siunitx}
\restoresymbol{SIX}{tablenum}

\received{July 22, 2022}
\revised{August 9, 2022}
\accepted{August 12, 2022}
\shorttitle{Serendipitous Detection of an \oiii\,$\lambda$5007 and \ha\ Emitter at $z=6.11$}
\shortauthors{Sun et al.}
\newcommand{\oii}{\mbox{[\ion{O}{2}]}}
\newcommand{\oiii}{\mbox{[\ion{O}{3}]}}
\newcommand{\heii}{\mbox{\ion{He}{2}}}
\newcommand{\nii}{\mbox{[\ion{N}{2}]}}
\newcommand{\hb}{\mbox{H$\beta$}}
\newcommand{\ha}{\mbox{H$\alpha$}}
\newcommand{\popiii}{\mbox{\ion{Pop}{3}}}
\newcommand\msun{\mbox{\si{M_\odot}}}

\newcommand\smpy{\mbox{\si{M_\odot.yr^{-1}}}}

\renewcommand{\micron}{\si{\micro\meter}}

\newcommand{\textred}[1]{{#1}}

\begin{document}


\title{
First Peek with JWST/NIRCam Wide-Field Slitless Spectroscopy: \\
Serendipitous Discovery of a Strong [\ion{O}{3}]/H$\alpha$ Emitter at \mbox{\boldmath $z=6.11$} 
}

\correspondingauthor{Fengwu Sun}
\email{fengwusun@arizona.edu}

\author[0000-0002-4622-6617]{Fengwu Sun}
\affiliation{Steward Observatory, University of Arizona, 933 N. Cherry Avenue, Tucson, AZ 85721, USA}

\author[0000-0003-1344-9475]{Eiichi Egami}
\affiliation{Steward Observatory, University of Arizona, 933 N. Cherry Avenue, Tucson, AZ 85721, USA}

\author[0000-0003-3382-5941]{Nor Pirzkal}
\affiliation{ESA/AURA STScI, 3700 San Martin Dr., Baltimore, MD, 21218, USA}

\author[0000-0002-7893-6170]{Marcia Rieke}
\affiliation{Steward Observatory, University of Arizona, 933 N. Cherry Avenue, Tucson, AZ 85721, USA}

\author[0000-0003-4850-9589]{Martha Boyer}
\affiliation{Space Telescope Science Institute, 3700 San Martin Drive, Baltimore, MD 21218, USA}

\author[0000-0001-6464-3257]{Matteo Correnti}
\affiliation{Space Telescope Science Institute, 3700 San Martin Drive, Baltimore, MD 21218, USA}

\author[0000-0002-5581-2896]{Mario Gennaro}
\affiliation{Space Telescope Science Institute, 3700 San Martin Drive, Baltimore, MD 21218, USA}

\author[0000-0001-8627-0404]{Julien Girard}
\affiliation{Space Telescope Science Institute, 3700 San Martin Drive, Baltimore, MD 21218, USA}

\author[0000-0002-8963-8056]{Thomas P. Greene}
\affiliation{Space Science and Astrobiology Division, NASA Ames Research Center, MS 245-6, Moffett Field, CA 94035 USA}

\author{Doug Kelly}
\affiliation{Steward Observatory, University of Arizona, 933 N. Cherry Avenue, Tucson, AZ 85721, USA}

\author[0000-0002-6610-2048]{{Anton M. Koekemoer}}
\affiliation{Space Telescope Science Institute, 3700 San Martin Drive, Baltimore, MD 21218, USA}

\author[0000-0002-0834-6140]{Jarron Leisenring}
\affiliation{Steward Observatory, University of Arizona, 933 N. Cherry Avenue, Tucson, AZ 85721, USA}

\author{Karl Misselt}
\affiliation{Steward Observatory, University of Arizona, 933 N. Cherry Avenue, Tucson, AZ 85721, USA}

\author[0000-0002-6500-3574]{Nikolay Nikolov}
\affiliation{Space Telescope Science Institute, 3700 San Martin Drive, Baltimore, MD 21218, USA}

\author[0000-0002-6730-5410]{Thomas L. Roellig}
\affiliation{NASA Ames Research Center, MS 245-6, Moffett Field, CA 94035, USA}

\author[0000-0003-2434-5225]{John Stansberry}
\affiliation{Space Telescope Science Institute, 3700 San Martin Drive, Baltimore, MD 21218, USA}

\author[0000-0003-2919-7495]{Christina C. Williams}
\affiliation{Steward Observatory, University of Arizona, 933 N. Cherry Avenue, Tucson, AZ 85721, USA}

\author[0000-0001-9262-9997]{Christopher N. A. Willmer}
\affiliation{Steward Observatory, University of Arizona, 933 N. Cherry Avenue, Tucson, AZ 85721, USA}

\collaboration{18}{(Members of the JWST/NIRCam Commissioning Team)}



\begin{abstract}
We report the serendipitous discovery of an \oiii\ $\lambda\lambda$4959/5007 and \ha\ line emitter in the Epoch of Reionization (EoR) with the JWST commissioning data taken in the NIRCam wide field slitless spectroscopy (WFSS) mode.
Located $\sim$55\arcsec\ away from the flux calibrator P330-E, this galaxy exhibits bright \oiii\ $\lambda\lambda$4959/5007 and \ha\ lines detected at 3.7, 9.9 and 5.7$\sigma$, respectively, with a spectroscopic redshift of $z=6.112\pm0.001$.
The total \hb+\oiii\ equivalent width is 664$\pm$98\,\AA\ \textred{(454$\pm$78\,\AA\ from the \oiii\,$\lambda$5007 line)}. 
This provides direct spectroscopic evidence for the presence of strong rest-frame optical lines (\hb+\oiii\ and \ha) in EoR galaxies as inferred previously from the analyses of Spitzer/IRAC spectral energy distributions.
Two spatial and velocity components are identified in this source, possibly indicating that this system is undergoing a major merger, which might have triggered the ongoing starburst with strong nebular emission lines over a timescale of $\sim$2\,Myr as our SED modeling suggests.
The tentative detection of \heii\,$\lambda$4686 line \textred{($1.9\sigma$)}, if real, may indicate the existence of very young and metal-poor star-forming regions with a hard UV radiation field.
Finally, this discovery demonstrates the power and readiness of the JWST/NIRCam WFSS mode, and marks the beginning of a new era for extragalactic astronomy, in which EoR galaxies can be routinely discovered via blind slitless spectroscopy through the detection of rest-frame optical emission lines.
\end{abstract}



\section{Introduction} \label{sec:01_intro}

In the Epoch of Reionization (EoR; $z\gtrsim 6$), the rest-frame optical nebular emission lines (e.g., \ha, \hb\ and \oiii\,$\lambda\lambda$4959/5007) of galaxies were difficult to observe in the pre-JWST era.
Typically being the brightest lines in the rest-frame optical bands, \oiii\ and \ha\ get redshifted out of the near-infrared (NIR) $K$ band at $z > 3.6$ and 2.5, respectively, and therefore become inaccessible from the ground because of low sensitivity.
At higher redshifts, these lines enter the passbands of the Spitzer/IRAC Channel 1/2 (3.6/4.5\,\micron), boosting broad-band photometric measurements \citep[e.g.,][]{schaerer09}.
Such an effect has been widely used to infer their line strengths at $z\simeq3.8 - 5$ \citep[\ha; e.g.,][]{shim11,stark13}, $z\simeq 5.1 - 5.4$ \citep[\ha; e.g.,][]{rasappu16} and $z\simeq 6.7 - 7.0$ \citep[\hb\ and \oiii; e.g.,][]{smit14,smit15,endsley21b,endsley21a}.
Possible modulation of observed IRAC [3.6\,\micron]--[4.5\,\micron] colors by emission lines was also reported for galaxies even at $z \sim 8$ \citep[e.g.,][]{labbe13,debarros19}.
Despite all this exciting evidence, up to now, all of these optical emission line studies in the EoR relied on broad-band photometry, which is inevitably affected by multiple factors such as the assumptions of star-formation history, metallicity, ionization parameter and dust attenuation. 

The James Webb Space Telescope (JWST) will undoubtedly revolutionize the studies of rest-frame optical-line emitters at high redshift (see a recent review by \citealt{robertson22}).
With its unprecedented sensitivity, JWST will detect \ha\ emission up to $z \sim 6.7$ and \oiii\ emission up to $z \sim 9$ with two of the four scientific instruments, NIRCam and NIRSpec.
The wide-field slitless spectroscopy mode of NIRCam \citep{greene17} conducts grism spectroscopy at 2.4--5.0\,\micron\ with a medium spectral resolution of $R \sim 1600$ around 4\,\micron. 
\textred{With a dispersion around 1\,nm/pixel, one cannot obtain the full 2.4--5\,\micron\ spectrum simultaneously across the detector which has 2040 pixels in each dimension. 
Therefore, a filter is also needed to ensure that desired wavelengths are captured on the detector. 
A filter will also minimize IR background and its associated noise from the telescope and sky.}
Combining the data taken with the two long-wavelength modules of NIRCam, an instantaneous survey area up to $\sim$9\,arcmin$^2$ can be reached \textred{(in which sources yield either partial or complete spectra)}, enabling efficient wide-field surveys of line-emitting galaxies in the EoR.
Multiple JWST Cycle-1 Early Release Science (ERS) and General Observers (GO) programs (e.g., CEERS, PI: Finkelstein; ASPIRE, PI: Wang; FRESCO, PI: Oesch) will exploit this powerful observing mode by detecting $>10^2 \sim 10^3$ \oiii\ and \ha\ line emitters at $z \gtrsim 6$.

In this work, we present the first serendipitous spectroscopic discovery of an \oiii\,$\lambda\lambda$4959/5007 and \ha\ line emitter at $z > 6$ with the JWST/NIRCam WFSS mode.
The galaxy, named NRCJ1631+3008-z6.1, was discovered in the field around the flux calibrator P330-E (GSC\,02581-02323), which was observed during the commissioning phase of this observing mode. The detection of \oiii\,$\lambda\lambda$4959/5007 and \ha\ emission lines at 3.528, 3.562 and 4.668\,\micron\ indicates a spectroscopic redshift of $z = 6.112\pm0.001$.
The observations and corresponding data reduction techniques are described in Section~\ref{sec:02_obs}.
The spectroscopic, photometric measurements and spectral energy distribution (SED) modeling are presented in Section~\ref{sec:03_res}.
In Section~\ref{sec:04_dis}, we compare our results with pre-JWST-era studies at $z\simeq 0 - 7$ and discuss the implication of emission-line strengths.
The conclusions can be found in Section~\ref{sec:05_con}.
Throughout this Letter, we assume a flat $\Lambda$CDM cosmology with $h = 0.7$ and $\Omega_m = 0.3$, and a \citet{chabrier03} initial mass function (IMF).
The AB magnitude system \citep{oke83} is used.

\begin{figure*}[!t]
\centering
\includegraphics[width=\linewidth]{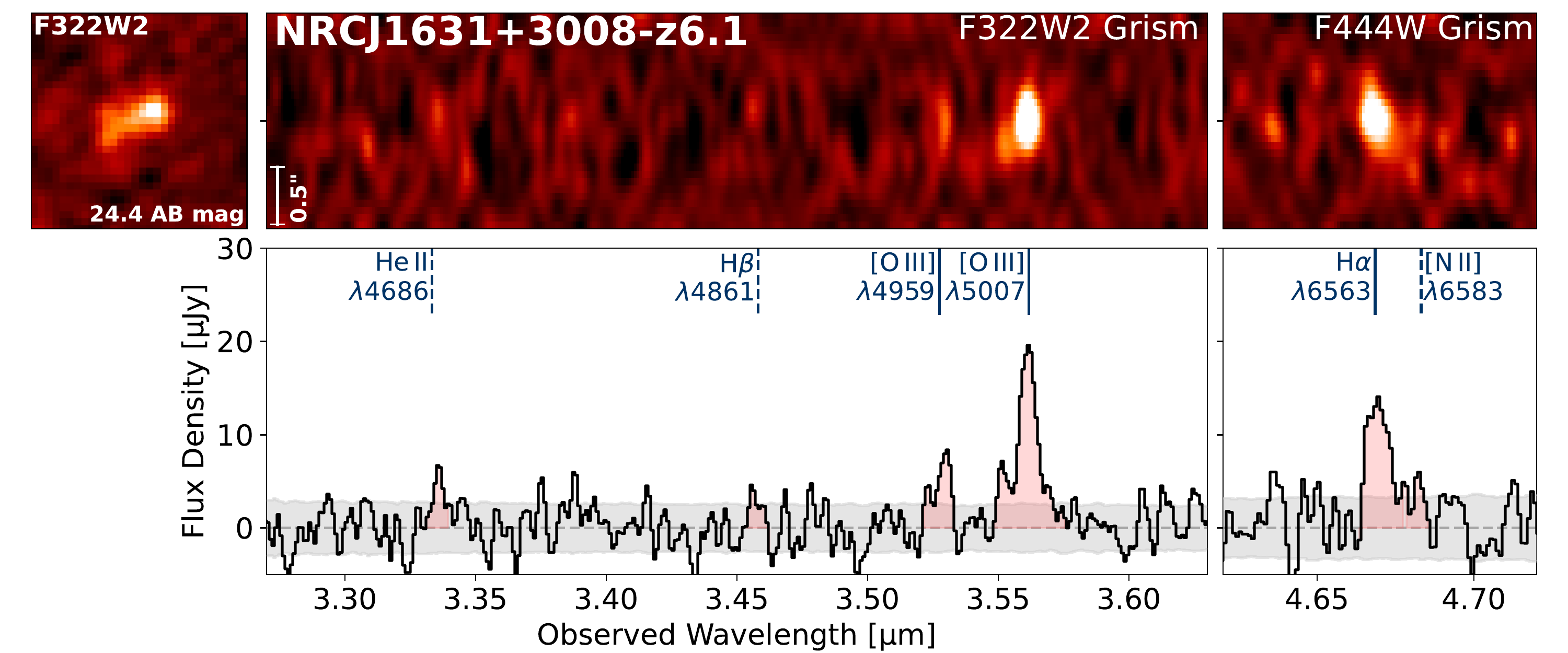}
\caption{JWST/NIRCam F322W2 image (top-left; size: 1\farcs9$\times$1\farcs9), 2D and 1D spectra (top/bottom-middle/right) of NRCJ1631+3008-z6.1 in the F322W2 and F444W band. 
Image and 1D spectra were smoothed with Gaussian kernel of $\sigma=1$\,pix (0\farcs063 or 1\,nm) and 2D spectra were smoothed with $\sigma=2$\,pix.
All spectra were flux-calibrated and background-subtracted.
1D spectra were extracted within $D=7$\,pix (0\farcs44) box with aperture correction factors applied (modeled from the profiles of \oiii\,$\lambda$5007 and \ha\ line).
Six identified lines (\heii\,$\lambda$4686, \hb, \oiii\,$\lambda\lambda$4959/5007, \ha\ and \nii\,$\lambda$6583) at $z=6.112\pm0.001$ are labeled with dark blue lines and highlighted with shallow red shades.
1D error spectra are shown as the gray-shaded regions.
}
\label{fig:01_spec}
\end{figure*}

\section{Observation and Data Reduction}
\label{sec:02_obs}

The JWST/NIRCam long-wavelength (LW; 2.4--5.0\,\micron) grism characterization observations were obtained through Program \#1076 (PI: Pirzkal) during the commissioning phase of the instrument.
NIRCam has four grisms: each of the two modules (A and B) has two grisms that disperse light parallel to
detector rows (Grism R) or columns (Grism C), respectively.
For the flux and wavelength calibration, P330-E (a G2V star) and IRAS\,05248-7007 (a post-Asymptotic-Giant-Branch star in the Large Magellanic Cloud) were observed, respectively.

The WFSS observations of the P330-E field were obtained with both F322W2 and F444W filters on April 29, 2022.
In each band, the target was observed with four grisms \textred{(AR, AC, BR, BC)} at four \textsc{intramodulex} primary dither positions, respectively.
With a five-group \textsc{bright1} readout pattern, the effective time per integration was 96\,s.
This resulted in a signal-to-noise ratio (S/N) of $\gtrsim$\,100 (per pixel, $\sim$1\,nm) for the extracted 1D spectrum of P330-E in each exposure, enabling accurate flux calibration over the full wavelength range of NIRCam WFSS mode.
We note that the last exposure with the Module B Grism C in the F322W2 band (Observation \#108, Visit 001, Exposure 4) failed because of unstable guiding, and therefore these data were not included in the analysis. 
The effective total exposure time of each source depends on its location, with a maximum of $\sim$25\,min in each band after combining exposures with all grisms.

The WFSS data were reduced to the level of Stage-1 (i.e., ``\textsc{\_rate}" files) with the standard \textsc{jwst} calibration pipeline\footnote{\href{https://github.com/spacetelescope/jwst}{https://github.com/spacetelescope/jwst}} v1.4.7.
After that, we performed the 2D sky-background subtraction using the sigma-clipped median images, which were constructed from the obtained WFSS data.
We then applied flat-field correction using the imaging flat data obtained with the same filter and module.
The world coordinate system (WCS) of each grism image was calibrated with the \textsc{Gaia} DR2 catalog \citep{gaiadr2} by matching with the stars detected in the NIRCam short-wavelength (SW) images, which were taken simultaneously in the F212N band.

The imaging data of the P330-E field were taken either simultaneously with the grism exposures (F212N) or after the grism observations through direct and out-of-field imaging (SW: F212N; LW: F250M, F322W2 and F444W).
The direct imaging data were reduced and mosaicked using the standard Stage-1/2/3 \textsc{jwst} pipeline.
We also manually adjusted the photometric zeropoints of these image products based on the in-flight measurements of P330-E (see \citealt{rigby22}) by $-$0.11, 0.06, 0.15 and 0.21 mag with the F212N, F250M, F322W2 and F444W bands, respectively, as the revised photometric zeropoints were not incorporated in the early pipeline version that we used.
\textred{The revised zeropoints are consistent with those in the latest reference files \texttt{`jwst\_0942.pmap'} at the time of writing.}
The final image products were resampled to a native pixel size of 0\farcs0312 (SW) and 0\farcs0629 (LW) with \texttt{pixfrac}\,$=$\,0.8,
and the WCS of the images were registered with the \textsc{Gaia} DR2 catalog \citep{gaiadr2}.  The effective exposure times ($t_\mathrm{exp}$) at the location of our source, NRCJ1631+3008-z6.1, are 33.8, 6.4, 2.5 and 3.2\,min in the F212N, F250M, F322W2 and F444W bands, respectively.

We also note that although P330-E was frequently observed as a flux calibrator for HST and JWST, most of the existing data are too shallow and our source was not detected.
Ground-based $g/r/z$-band images (e.g., the Mosaic $z$-band Legacy
Survey and the Beijing-Arizona Sky Survey; \citealt{dey19}, \citealt{zou19}) were also checked, but the data were too shallow to constrain the SED ($3\sigma$ upper limits at $g>24.8$ and $z>23.6$\,mag) of NRCJ1631+3008-z6.1.

We performed source extraction with the mosaicked F322W2 and F444W images using \textsc{SExtractor} v2.25.3 \citep{1996A&AS..117..393B}.
With the derived source catalog, we conducted 2D spectral extraction, wavelength and flux calibration on the flat-fielded WFSS data, and combined the extracted data taken with Grism R, C or both.
These were performed for $\sim$3000 sources detected in the F322W2 image, to test the accuracy of derived spectral tracing and dispersion models in the instrument commissioning phase. 
We defer the detailed descriptions of these models to another paper from the collaboration.
Qualitatively, for a given source with direct-imaging position of ($x_0$, $y_0$) in an exposure, our dispersion model can predict the position of a spectral feature at a wavelength of $\lambda_{s}$ along the dispersion direction (e.g., $x_{s}$ in Grism R) with a root-mean-square (RMS) accuracy of 0.2\,pixel, i.e., $\sim$\,10\%\ of the resolution element.
With the derived $x_{s}$, our spectral tracing model can predict the position along the perpendicular direction (e.g., $y_s$ in Grism R) with an RMS accuracy of 0.1--0.2\,pixel, i.e., 5--10\%\ of the full width at half maximum (FWHM) of the point-source spectral trace.

\section{Results}
\label{sec:03_res}

\subsection{Discovery of an \oiii\,$\lambda$5007 emitter at $z=6.112$}\label{ss:03a_disc}

Through visual inspection of extracted spectra, we identified a source with a prominent emission line feature at 3.562\,\micron\ (NRCJ1631+3008-z6.1; Figure~\ref{fig:01_spec}). 
Located $\sim$55\arcsec\ away from P330-E, this source was observed with Grism R in five integrations of exposure ($t_\mathrm{exp} = 8$\,min) and Grism C in seven integrations ($t_\mathrm{exp} = 11$\,min).
We note that this source could potentially yield spectra in two additional integrations with Module A Grism R (Observation \#106, Visit 001, Exposure 2/3), but the direct-imaging position of the source was near the edge of the pick-off mirror. 
We did not include these data as the emission line was not seen in these integrations, suggesting a significant loss of the target signal.
The 2D spectra of this source overlapped with two continuum sources around 3.5\,\micron\ in Module B Grism R (four integrations).
In the combined 2D spectra, we subtracted 1D continuum and background along both the row and column direction through linear fitting in line/source-free region. \textred{This reduced the RMS noise by $\sim$20\% and the resultant background can be well described with a normal distribution with zero mean through 
Kolmogorov–Smirnov test ($p$-value=0.43)}.
1D spectra were extracted using a box aperture with height $D=7$\,pix (0\farcs44), which optimized the S/N of the second brightest emission line at 3.528\,\micron.
An aperture correction factor was calculated as 1.07 using the 3.562-\micron\ line profile in the vertical direction.

\begin{figure*}[!t]
\centering
\includegraphics[width=\linewidth]{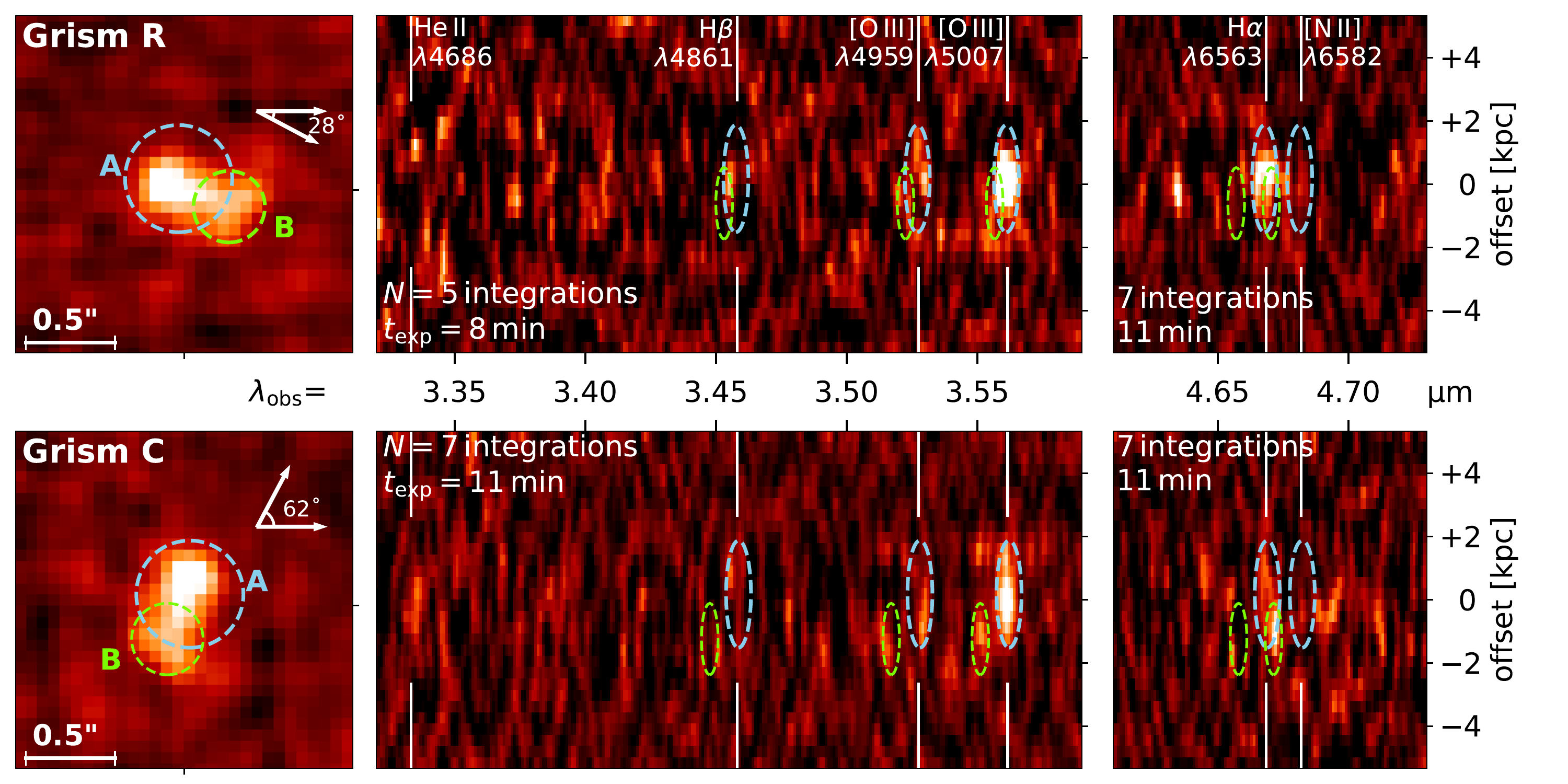}
\caption{JWST/NIRCam F322W2 direct image and F322W2/F444W 2D spectra of NRCJ1631+3008-z6.1 seen with Grism R (top) and C (bottom).
The orientations of direct images have been aligned to the dispersion directions of the two grisms.
In the plots of 2D spectra, the expected positions of detectable emission lines are indicated with vertical white lines. 
\oiii\,$\lambda\lambda$4959/5007 and \ha\ lines can be detected with Grism R and C data separately, confirming the line identification.
Two potential clumpy structures (A and B) with spatial offset of 1.6$\pm$0.3\,kpc and velocity offset of 602$\pm$105\,\si{km.s^{-1}}  are noted with dashed blue and green circles in direct images, respectively, and the expected locations of corresponding line emissions (\oiii, \ha\ and \nii) are shown with dashed circles of the same colors superimposed on the 2D spectra.
The PAs of the two clumps with respect to the dispersion directions are also shown in the direct images.
}
\label{fig:02_RC}
\end{figure*}

The two lines detected at 3.562\,\micron\ (9.9$\sigma$) and 3.528\,\micron\ (3.7$\sigma$) were identified as \oiii\,$\lambda\lambda$5007/4959 emissions at $z=6.112\pm0.001$.
We also searched for other possible emission lines in extracted spectra, and tentatively identified \heii\,$\lambda$4686 (1.9$\sigma$) and \hb\,$\lambda$4861 (1.5$\sigma$) lines.
\oii\,$\lambda\lambda$3726/3729 doublet lines were not detected in the F322W2 spectrum ($<4\times10^{-17}$\,\si{erg.s^{-1}.cm^{-2}}) because of a twice higher RMS noise at the expected wavelength.
We also identified \ha\ (5.7$\sigma$) and \nii\,$\lambda$6583\ (1.4$\sigma$) in the F444W spectra at the expected wavelengths (right panels of Figure~\ref{fig:01_spec}).
We modeled the profiles of four identified lines in F322W2 band with Gaussian functions simultaneously, and in this fitting, the FWHM and flux of each line were allowed to float, while the line centers were controlled by the redshift parameter.
The best-fit spectroscopic properties are presented in Table~\ref{tab:01}.
Measurements using grism R-only or C-only 1D spectra returned consistent line fluxes.
We observed a flux ratio of $2.9\pm0.8$ between the identified \oiii\,$\lambda\lambda$5007/4959 lines, consistent with the theoretical ratio of three. 
\ha\ and \nii\ lines were modeled with similar methods, and the detailed analysis of these lines will be presented in another paper from the collaboration with a larger sample of $z\gtrsim6$ \ha\ emitters.


\begin{table}[!t]
\caption{Summary of the properties of NRCJ1631+3008-z6.1.}
\label{tab:01}
\small
\renewcommand{\arraystretch}{0.9}
\begin{tabular}{lr}
\hline\hline Photometric Properties                        \\\hline
R.A.                      & 16:31:34.45                    \\
Decl.                     & +30:08:10.6                    \\
F212N [AB\,mag]           & $>$23.8                        \\
F250M [AB\,mag]           & $>$24.8                        \\
F322W2 [AB\,mag]          & 24.45$\pm$0.09                 \\
F444W [AB\,mag]           & 24.53$\pm$0.17                 \\
\hline\hline Spectroscopic Properties                      \\\hline
Redshift                  & 6.112$\pm$0.001                \\
$f$(\heii\,$\lambda$4686) [\si{10^{-18}\,erg.s^{-1}.cm^{-2}}] & 9.9$\pm$5.2                    \\
$f$(\hb) [\si{10^{-18}\,erg.s^{-1}.cm^{-2}}]            & 5.0$\pm$3.2                    \\
$f$(\oiii\,$\lambda$4959) [\si{10^{-18}\,erg.s^{-1}.cm^{-2}}] & 14.2$\pm$3.9                   \\
$f$(\oiii\,$\lambda$5007) [\si{10^{-18}\,erg.s^{-1}.cm^{-2}}] & 40.5$\pm$4.1                   \\
$f$(\ha) [\si{10^{-18}\,erg.s^{-1}.cm^{-2}}]            & 16.0$\pm$2.8                   \\
$f$(\nii\,$\lambda$6583) [\si{10^{-18}\,erg.s^{-1}.cm^{-2}}] & 3.3$\pm$2.3                    \\
EW(\heii\,$\lambda$4686) [\AA] & 97$\pm$52                      \\
EW(\hb) [\AA]             & 52$\pm$35                      \\
EW(\oiii\,$\lambda$4959) [\AA] & 156$\pm$47                     \\
EW(\oiii\,$\lambda$5007) [\AA] & 454$\pm$78                     \\
EW(\ha) [\AA]             & 369$\pm$104                    \\
EW(\nii\,$\lambda$6583) [\AA] & 76$\pm$56                      \\
EW(\hb$+$\oiii) [\AA]     & 664$\pm$98                     \\
FWHM(\oiii\,$\lambda$5007) [\si{km.s^{-1}}] & 382$\pm$130                    \\
FWHM(\ha) [\si{km.s^{-1}}]               & 267$\pm$109                    \\
\hline\hline Physical Properties                           \\\hline
$R_\mathrm{e}$ [kpc]      & 1.2$\pm$0.1                    \\
SFR(\ha) [\smpy]                         & 36$\pm$6                       \\
SFR(UV,SED) [\smpy]                      & 23$\pm$6                       \\
$M_\mathrm{star}$ [$10^8$\,\msun]        & 10$\pm$8                       \\
$\log[\xi_\mathrm{ion}/(\mathrm{erg}^{-1}\mathrm{Hz})]$ & 25.3$\pm$0.1                   \\
\hline
\end{tabular}
\tablecomments{\textred{The uncertainty of absolute flux calibration is estimated as between 2\%  (according to \citealt{gordon22}) and 5\% (conservative), considerably smaller than the observational errors presented in this table.}
\textred{The derivation of EWs is detailed in Section~\ref{ss:03c_sed}.}
}
\renewcommand{\arraystretch}{1.0}
\end{table}

\subsection{Two spatial and velocity components}
\label{ss:03b_dual}

Figure~\ref{fig:02_RC} displays the 2D spectra of NRCJ1631+3008-z6.1 obtained with the orthogonal Grism R and C separately. 
\oiii\ and \ha\ lines can be detected in the two spectra individually, further confirming the line identification and association with the source.

From the F322W2 image, we identified two spatial components labeled as clump-A and B with a physical separation of $1.6\pm0.3$\,kpc between their centroids.
The dispersion direction of Grism C was off by 62$\pm$7\arcdeg\ from the position angle (PA) of the two clumps, and two clumps of \oiii\,$\lambda$5007\ emissions can be identified on the 2D spectrum with Grism C.
Similar clumps can also be tentatively seen for \oiii\,$\lambda$4959, albeit at much lower significance.
These can also been seen as the major and minor peaks of \oiii\ lines in the 1D spectrum (Figure~\ref{fig:01_spec}).
Through both 1D line profile fitting and aperture photometry of the \oiii\,$\lambda$5007 emission lines in the spectra, we measured a flux ratio of 2.6$\pm$0.6 between the two \oiii\ clumps. 

The offset between the two clumps in spectral image is a result of both intrinsic velocity offset and projected physical offset along the dispersion direction, and the velocity offset was measured as 605$\pm$102\,\si{km.s^{-1}}. 
In the Grism R data, the dispersion direction was only off by 28$\pm$7\arcdeg\ from the PA between clump-A/B, leading to a strong spectral overlapping between the two clumps.
As a result, the overall line width seen with Grism R, which was convolved with the surface brightness profile of the source in dispersion direction, is also larger than that measured from the Grism C data.
The deconvolved (intrinsic) FWHM of \oiii\,$\lambda$5007 line is measured as $362\pm134$\,km/s from both the Grism R and C data.

We also examined such a two-component model with the F444W spectra.
Interestingly, we found that the bulk of the \ha\ emission is from clump-A, which can also be tentatively seen for \hb\ emission with F322W2 data despite a lower significance.
In contrast, clump-B hosts more luminous \nii\ emission that overlaps with the \ha\ emissions from clump-A in the wavelength space with both grisms.
The \nii\ emissions from clump-B can be clearly seen in Grism C data where the perpendicular offset of the two clumps is larger. 
A higher \nii/\ha\ ratio \textred{in clump-B ($>$\,1.2) compared with clump-A ($0.25 \pm 0.17$)} may indicate elevated metallicity and/or harder UV radiation field in clump-B.

\begin{figure}[!t]
\centering
\includegraphics[width=\linewidth]{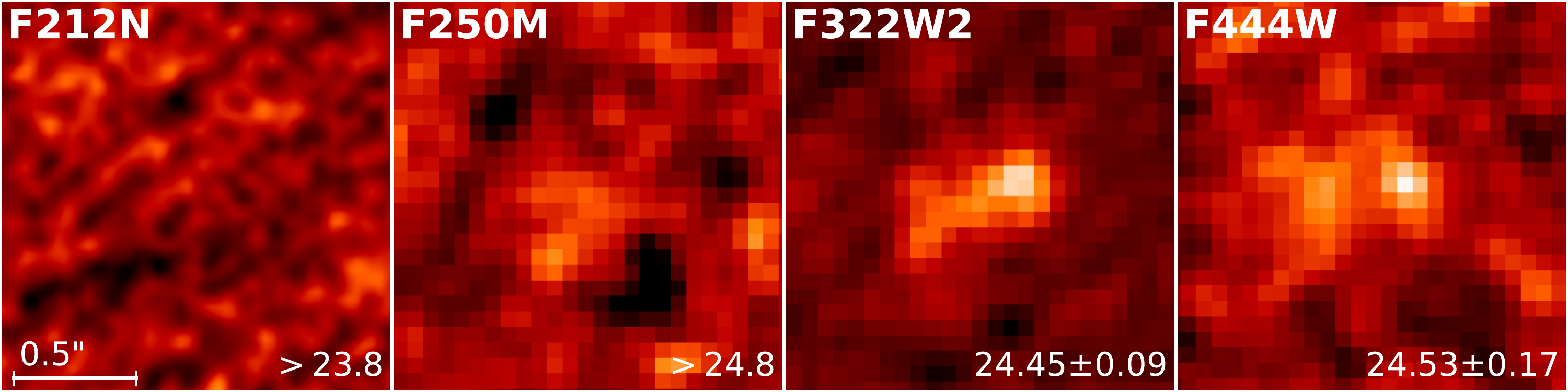}
\includegraphics[width=\linewidth]{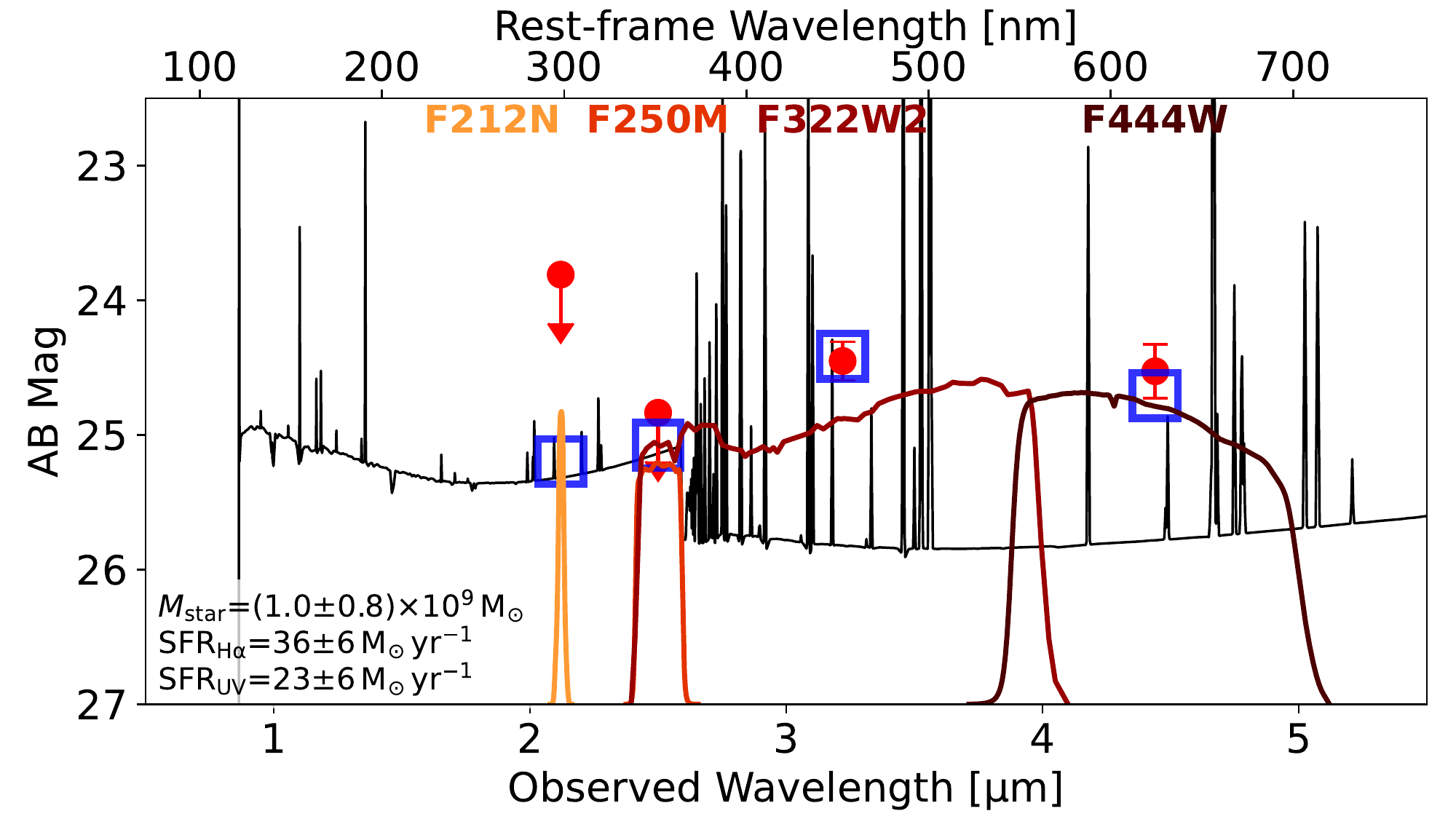}
\caption{Top: JWST/NIRCam short-wavelength (F212N) and long-wavelength (F250M, F322W2 and F444W) cutout images of NRCJ1631-3008-z6.1. 
Measured source brightness in each filter is shown in the lower-right corner of each plot (in AB\,mag; $3\sigma$ upper limit for non-detection).
Bottom: Best-fit SED model obtained with \textsc{cigale} \citep[black curve;][]{boquien19}.
Photometric measurements are shown as red circles, and best-fit source brightnesses are shown as open blue squares.
The transmission curves of four utilized filters are also plotted for comparison.
Note that the transmission of F322W2 band is computed using the measured grism throughput and is different from that in the JWST User Documentation. 
}
\label{fig:03_sed}
\end{figure}

\subsection{Photometry and SED modeling}
\label{ss:03c_sed}

We performed aperture photometry of NRCJ1631+3008-z6.1 (both clumps together) in F212N, F250M, F322W2 and F444W images with a circular aperture of $r=0\farcs45$ using \textsc{Photutils} \citep{photutils}.
The circularized surface brightness profile modeled in the F322W2 band (effective radius $R_\mathrm{e}=0\farcs22\pm0\farcs02$; 1.2$\pm$0.1\,kpc after deconvolving the point-spread function) suggests that the aperture correction is negligible. 
The sky background was subtracted using the median of sigma-clipped local annulus, and the photometric uncertainty was computed using the RMS of that.
The source was detected in the wide F322W2 (24.45$\pm$0.09 mag) and F444W (24.53$\pm$0.17 mag) bands, but remained undetected in either the F212N or F250M band ($3\sigma$ upper limits of $>$23.8 and $>$24.8 mag, respectively; Figure~\ref{fig:03_sed}).
\textred{The non-detection in the F212N/F250M band can be explained by their smaller bandwidths in contrast to those of wide filters, and lack of luminous line emission within their wavelength ranges.}

We then conducted SED modeling with \textsc{cigale} \citep{boquien19}.
In addition to the four-band photometry, the equivalent widths (EWs) of \hb, \oiii\ and \ha\ lines were also included as constraints.
These line EWs were calculated using the measured line fluxes and underlying F322W2/F444W continuum flux densities,
which were computed from broad-band photometry by subtracting all measured line fluxes within the passband and were assumed as constant across the bandwidth. 
All line EWs are reported in Table~\ref{tab:01}.
We assumed a commonly used delayed star-formation history (SFH; \texttt{sfhdelayed}) with an optional late starburst, and allowed a metallicity range of 0.2\,\si{Z_{\odot}}$\,\sim$\,\si{Z_{\odot}}.
The allowed range of ionization parameter ($\log U $) was set as $-1 \sim -2$ given the presence of strong \oiii\ and \heii\ lines, and a modified \citet{calzetti00} attenuation curve was adopted.

The best-fit SED model is shown in the lower panel of Figure~\ref{fig:03_sed}. The best-fit SFH model invokes both a young (1--2\,Myr) and old ($\sim$500\,Myr) stellar populations to interpret both strong optical emission lines and underlying stellar continuum.  
Similar models have been reported for $z>6$ galaxies with the presence or evidence of strong nebular lines \citep[e.g.,][]{hashimoto18,jiang20,whitler22}.
\textred{The reduced $\chi^2$ of best-fit line EWs is 0.65.}
The dust attenuation is negligible, and the best-fit metallicity is $\sim$\,0.4\,\si{Z_\odot}.
We derived a stellar mass of $(1.0\pm0.8)\times10^{9}$\,\msun.
In order to compare our star-formation rate (SFR) measurement with rest-frame-UV-selected galaxies at $z>6$, we derived a UV SFR of 23$\pm$6\,\smpy\ from the best-fit SED.
Although the UV continuum of NRCJ1631+3008-z6.1 was not directly detected in any band, the inferred SFR(UV, SED) is comparable with the SFR derived with \ha\ luminosity (36$\pm$6\,\smpy) assuming the \ha\ luminosity--SFR conversion in \citet{ke12}.

\section{Discussion}
\label{sec:04_dis}

\begin{figure*}
\centering
\includegraphics[width=0.328\linewidth]{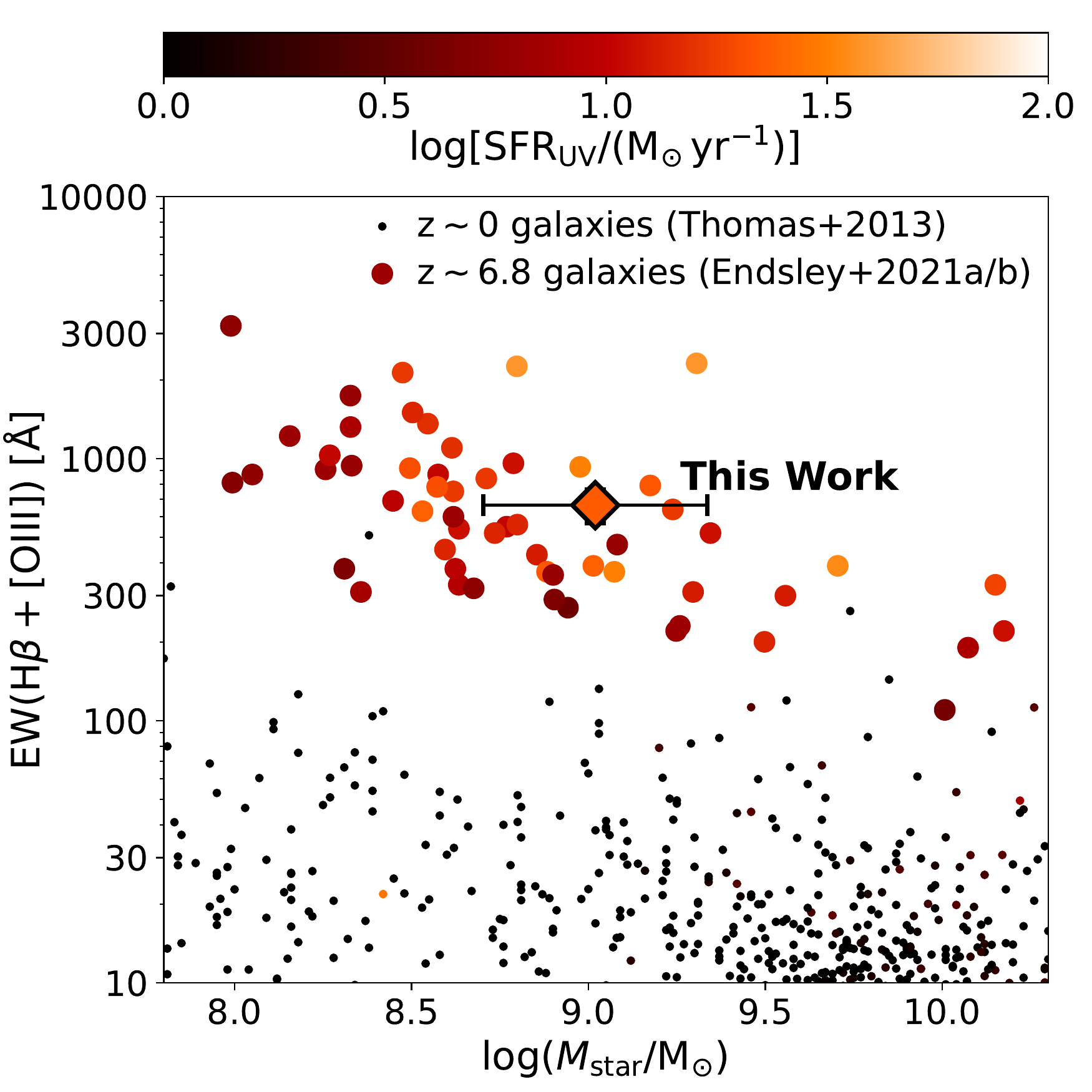}
\includegraphics[width=0.328\linewidth]{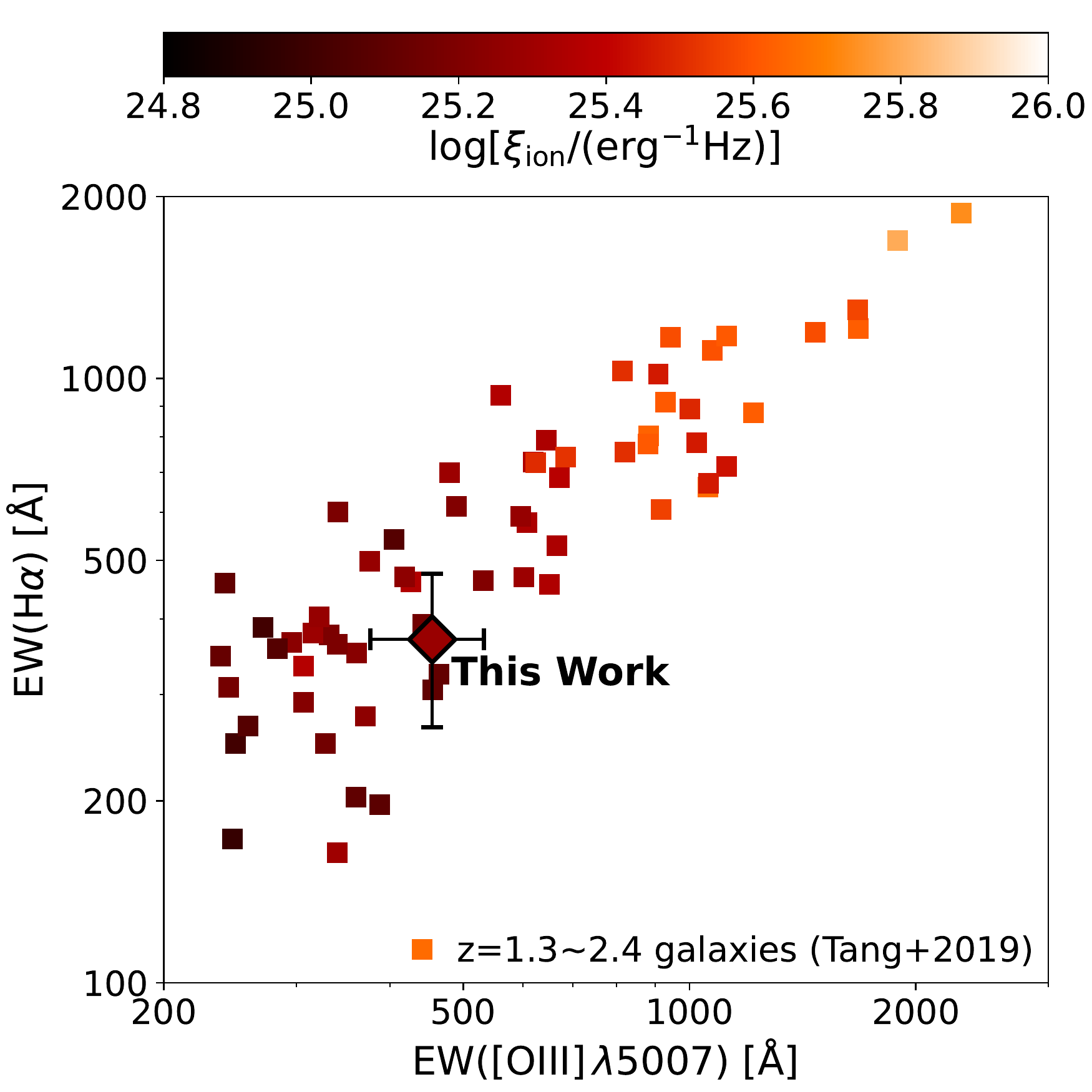}
\includegraphics[width=0.328\linewidth]{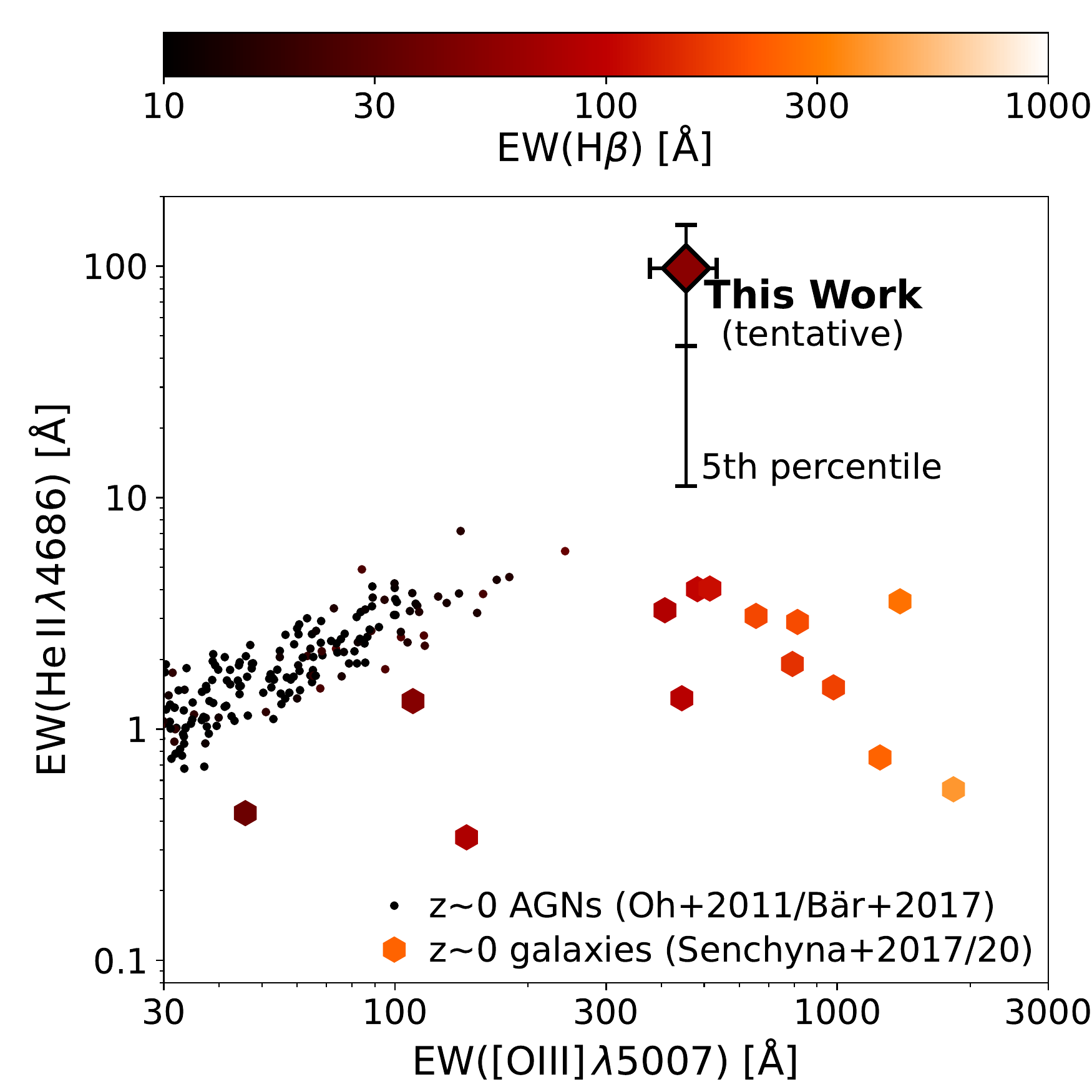}
\caption{Comparisons of emission line equivalent widths of NRCJ1631+3008-z6.1 (colored diamonds with black edges) with those of galaxies across $z\simeq 0 -7$.
Left: \hb$+$\oiii\ line EW versus stellar mass of NRCJ1631+3008-z6.1 compared with those of $z\sim6.8$ galaxies whose EWs are inferred from Spitzer/IRAC [3.6\,\micron]--[4.5\,\micron] color \citep[circles;][]{endsley21b,endsley21a}, and also SDSS-selected galaxies in the local Universe (dots; \citealt{thomas13}). 
Sources at $z>6$ are color-coded by their UV-based SFRs.
Middle: \ha\ EW versus \oiii\,$\lambda$5007\ EW compared with those of extreme emission-line galaxies at $z=1.3-2.4$ \citep[squares;][]{tang19}.
Sources are color-coded by their ionizing photon production efficiency ($\xi_\mathrm{ion}$).
Right: \heii\,$\lambda$4686\ EW versus \oiii\,$\lambda$5007\ EW compared with those of extreme emission-line galaxies  \citep[hexagons;][]{senchyna17,senchyna20} and SDSS-selected \heii\ AGNs \citep[dots;][]{oh11,bar17} in the local Universe.
Sources are color-coded with their \hb\ EWs.
Given the low significance of \heii\,$\lambda$4686 detection with NRCJ1631+3008-z6.1, we show the 5th, 16th and 84th percentiles of the likelihood distribution of the \heii\ EW.
}
\label{fig:04_ew}
\end{figure*}

\subsection{Comparisons with galaxies across $z\simeq 0 - 7$}
\label{ss:05a_compare}

We compare the emission-line and physical properties of NRCJ1631+3008-z6.1 with those of galaxies across $z\simeq 0 - 7$ in Figure~\ref{fig:04_ew}.

In the left panel, we first compare the measured \hb$+$\oiii\ EW with $z\sim6.8$ galaxies in \citet{endsley21b,endsley21a}.
In the pre-JWST era, \hb$+$\oiii\ EWs at $z\gtrsim4$ can only be inferred through the modeling of Spitzer/IRAC SEDs within certain redshift windows (e.g., $z\simeq 6.7 - 7.0$).
\citealt{endsley21a} reported a median \hb$+$\oiii\ EW of $759^{+112}_{-113}$\,\AA\ for UV-luminous ($M_\mathrm{UV}\lesssim-21$, $\mathrm{SFR}_\mathrm{UV} \gtrsim 10$\,\smpy) galaxies at $z \simeq 6.8$.
With the data presented here, we show that the \hb$+$\oiii\ EW of NRCJ1631+3008-z6.1 (664$\pm$98\,\AA) is consistent with those of $z\sim 6.8$ galaxies with comparable stellar mass and $\mathrm{SFR}_\mathrm{UV}$.
This confirms the presence of prominent rest-frame optical nebular emission lines in EoR galaxies, which are more than an order-of-magnitude stronger than those of galaxies in the local Universe with comparable stellar masses (e.g., SDSS-selected galaxies; \citealt{maraston13}, \citealt{thomas13}).

In the middle panel, we compare the \ha\ and \oiii\,$\lambda$5007 EWs of NRCJ1631+3008-z6.1 with extreme emission-line galaxies at $z=1.3-2.4$ \citep{tang19}, where these redshifted optical emission lines are still accessible for ground-based NIR spectroscopy.
The \ha\ EW of our source is consistent with those of $z=1.3-2.4$ galaxies with similar \oiii\ EWs.
We also compute the hydrogen ionizing photon production efficiency ($\xi_\mathrm{ion}$) using the observed \ha\ (tracing ionizing photons) and inferred UV luminosity following the same procedure in \citet{shivaei18} and \citet{tang19}.
We derive a $\log[\xi_\mathrm{ion} / (\mathrm{erg}^{-1}\mathrm{Hz})] = 25.3\pm0.1$ for NRCJ1631+3008-z6.1, which is widely seen in emission-line galaxies at $z\sim2$ \citep[e.g.,][]{shivaei18}.
The derived $\xi_\mathrm{ion}$ is also consistent with those of the $z\sim2$ galaxies on the $\xi_\mathrm{ion} - \mathrm{EW(\oiii)}$ relation modeled by \citet{tang19}.  
This suggests that the physical properties of certain strong line-emitting galaxies at $z\sim2$ and $z>6$ may be similar.

One potential surprising discovery of this work is the tentative detection of \heii\,$\lambda$4686 line. 
\textred{We examine the strength of the \heii\ line in all twelve available integrations. 
The line flux remains positive in nine integrations (75\%), consistent with the expectation (72\%) from Gaussian statistics for a $2\sigma$ stacked signal.}
The production of helium recombination lines require sufficient ionizing photons with energy greater than 54.4\,eV, much higher than the requirements for hydrogen (13.6\,eV) and \oiii\ (35.1\,eV) lines, and a very low metallicity environment ($\lesssim$\,0.01\,\si{Z_{\odot}}) is required to interpret a large \heii\ EW ($\gtrsim$\,100\,\AA; e.g., \citealt{inoue11}).
In the right panel of Figure~\ref{fig:04_ew}, we show the EW(\heii\,$\lambda$4684) versus EW(\oiii\,$\lambda$5007) of NRCJ1631+3008-z6.1 compared to those of extreme emission-line galaxies \citep[][mostly powered by X-ray binaries]{senchyna17,senchyna20} and SDSS active galactic nuclei \citep[AGNs;][]{oh11,bar17} in the local Universe, which were detected in \heii\,$\lambda$4686.
At a given \oiii\ EW, the \heii\ EW of NRCJ1631+3008 is higher than those of all comparison samples at $z\sim 0$.

We note that a well-known Lyman-$\alpha$-emitting galaxy, CR7 at $z=6.6$, was reported for a possible detection of \heii\,$\lambda$1640 lines with ground-based and HST grism spectroscopy (\citealt{sobral15,sobral19}; but see also \citealt{shibuya18} claiming no detection of this line).
\heii\,$\lambda$4686 line is 60\% more luminous than \heii\,$\lambda$1640 line in Case B recombination with electron temperature $T_e=10^4$\,K \citep{osterbrock06}.
The tentative detection of \heii\,$\lambda$4686\ line may indicate an elevation of the hardness of the ionizing spectrum in certain EoR galaxies, i.e., a larger fraction of ionizing photons with high energy ($>$\,54.4\,eV) when compared with those seen at ionizing environments at lower redshifts (\heii\ $\mathrm{EW}\lesssim 10$\,\AA\ even in so-called Wolf-Rayet galaxies; e.g., \citealt{brinchmann08}, \citealt{cassata13}).
This may further indicate a young (1--2\,Myr) metal-poor (i.e., ``\popiii''-like) stellar population produced with a top-heavy IMF \citep[e.g.,][]{jerabkova17} with the existence of very massive stars over 300\,\msun\ \citep[e.g.,][which could produce \heii\ EW at $\gtrsim$100\,\AA]{schaerer02} and/or a higher binary fraction.
However, we also make the caveat clear that the significance of \heii\ detection is low ($1.9\sigma$) with the shallow commissioning data and no decisive conclusion could be made. 
Deeper \heii\,$\lambda$1640 or $\lambda$4684 spectroscopy is necessary to disentangle the physics behind this puzzling tentative detection.

\subsection{Overall physical picture of NRCJ1631+3008-z6.1}
\label{ss:05b_picture}

The two spatial 
and velocity 
components of NRCJ1631+3008-z6.1 resolved with JWST/NIRCam imaging and slitless spectroscopy suggest that this source is a galaxy pair.
Assuming a constant mass-to-light ratio for the two components, the mass ratio is $2.6\pm0.6$, satisfying the classification criterion of a major merger event.
The dynamic timescale computed from the spatial and velocity offset of the two components is 2--3\,Myr, consistent with the timescale of the late starburst (1--2\,Myr) indicated by the strong nebular lines as modeled in Section~\ref{ss:03c_sed}.
Therefore, it is possible that the strong optical emission lines and ongoing star formation seen in this source is triggered by the merger of the two galaxies undergoing the final coalescence phase.

\textred{The two components are unlikely two star-forming clumps in the same galaxy, because the corresponding dynamic mass $M_\mathrm{dyn} = v^2 R / G \sim 10^{11}$\,\msun\ would be much larger than one typically expects for a galaxy with a stellar mass of $10^{9}$\,\msun.
In addition to this, the physical extent of NRCJ1631+3008-z6.1 is also twice larger than those of galaxies with comparable UV luminosity and stellar mass at $z\sim6$ \citep[e.g.,][]{shibuya15}, making the system more likely a merger pair.
}

Although the \oiii, \hb\ and \ha\ line properties of NRCJ1631+3008-z6.1 are consistent with those of emission-line galaxies at intermediate ($z\sim2$) and high redshifts ($z \sim 6.8$), the tentative \heii\ detection is not expected based on the majority of existing literature except for a few rare cases (e.g., CR7; \citealt{sobral19}).
We note that both CR7 and NRCJ1631+3008-z6.1 show clear morphological and kinematic evidence of merger \citep{sobral15,matthee17}, and it is possible that \heii\ can originate from certain merging young and gaseous clumps with very low metallicity, which are barely enriched before the coalescence. 
The spatial distribution of metallicity in such systems can be uneven as shown in the 2D spectra of the \nii\ line.
Given the shallowness of obtained data, deeper integral field unit observations (e.g., with JWST/NIRspec) will be highly valuable to shed light on the potential existence of low-metallicity star-forming clumps in galaxies at the so-called ``Cosmic Dawn'' ($z\gtrsim6$).

\section{Conclusion}
\label{sec:05_con}

We report the serendipitous discovery of an \oiii\,$\lambda\lambda$4959/5007 and \ha\ line-emitting galaxy in the Epoch of Reionization with JWST/NIRCam wide-field slitless spectroscopy.
The galaxy, NRCJ1631+3008-z6.1, is located $\sim$\,55\arcsec\ apart from the flux calibrator P330-E, and was observed with JWST/NIRCam in the F322W2 and F444W band using both the imaging and slitless spectroscopy modes during the commissioning phase.
\oiii\,$\lambda\lambda$4959/5007 and \ha\ lines were detected in both the Grism R and C data separately, confirming the line identification with a spectroscopic redshift of $z=6.112\pm0.001$.
\heii\,$\lambda$4686, \hb\ and \nii\,$\lambda$6583 lines are also detected tentatively.

The equivalent widths of \oiii\, \hb, and \ha\ lines are consistent with those of emission-line galaxies at $z\simeq1-2$, which are measured through ground-based near-IR spectroscopy, and also those of galaxies at $z\sim6.8$, which are inferred from Spitzer/IRAC [3.6\,\micron]--[4.5\,\micron] colors.
This provides direct spectroscopic evidence for the presence of strong rest-frame optical nebular emission lines in galaxies at $z>6$.
The tentative detection of \heii\,$\lambda$4686, however, is inconsistent with those observed in either extreme emission-line galaxies or AGNs in the local Universe. 
If real, this may indicate the existence of very young and metal-poor star-forming regions with a hard UV radiation field in NRCJ1631+3008-z6.1.

We show that NRCJ1631+3008-z6.1 consists of two spatial and velocity components (clump-A/B) that are undergoing the final coalescence stage of a major merger.
The dynamic timescale (2--3\,Myr) of merger is consistent with the timescale of the late starburst (1--2\,Myr) modeled through SED fitting, indicating that the merger triggers the recent starburst and strong optical nebular emission lines (\oiii\ and \ha).
Clump-B shows higher \nii/\ha\ ratio than clump-A, which may indicate a higher metallicity.
Under-enriched gaseous clumps can be the source of potential \heii\ emission seen in $z>6$ merging galaxies, which could be tested with deep near-IR integral field spectroscopy in the JWST era.

Finally, the serendipitous discovery of NRCJ1631+3008-z6.1 in the commissioning phase demonstrates the scientific readiness and excellent capability of JWST/NIRCam WFSS mode.
It also marks the beginning of the JWST era for extragalactic astronomy, in which galaxies at $z\gtrsim 6$ can be routinely discovered and confirmed through wide-field slitless spectroscopy (see also the detections of \oiii\,$\lambda\lambda\lambda$4363/4959/5007 lines for $z>6$ galaxies with JWST/NIRSpec early release observations; \citealt{pontoppidan22}).


\begin{acknowledgments}
We thank Mengtao Tang for sharing his measurements of emission-line galaxies at $z=1.3-2.4$.
\textred{We thank the anonymous referee for a helpful and prompt report.}
\textred{We thank Pavel Kroupa and Daniel Stark for helpful comments.}
FS, EE, MR, DK, JL, KM, CCW and CNAW acknowledge funding from JWST/NIRCam contract to the University of Arizona, NAS5-02105.
This work is based on observations made with the NASA/ESA/CSA James Webb Space Telescope. The data were obtained from the Mikulski Archive for Space Telescopes at the Space Telescope Science Institute, which is operated by the Association of Universities for Research in Astronomy, Inc., under NASA contract NAS 5-03127 for JWST. These observations are associated with program \#1076.
{This Letter is based upon High Performance Computing (HPC) resources supported by the University of Arizona TRIF, UITS, and Research, Innovation, and Impact (RII) and maintained by the UArizona Research Technologies department.}
{All of the data presented in this Letter were obtained from the Mikulski Archive for Space Telescopes (MAST) at the Space Telescope Science Institute. The specific observations analyzed can be accessed via \dataset[10.17909/f8p1-e696]{https://doi.org/10.17909/f8p1-e696}. }
\end{acknowledgments}

%

\facilities{JWST (NIRCam)}


\software{
\textsc{astropy} \citep{2013A&A...558A..33A,2018AJ....156..123A},  
\textsc{SExtractor} \citep{1996A&AS..117..393B},
\textsc{Photutils} \citep{photutils},
\textsc{cigale} \citep{boquien19}
}






\bibliography{00_main}{}
\bibliographystyle{aasjournal}



\end{document}